# Ballistic edge states in Bismuth nanowires revealed by SQUID interferometry

**Authors:** Anil Murani[1], Alik Kasumov[1,2], Shamashis Sengupta[1], Yu.A. Kasumov[2], V.T.Volkov[2], I.I. Khodos[2], F. Brisset[3], Raphaëlle Delagrange[1], Alexei Chepelianskii[1], Richard Deblock[1], Hélène Bouchiat[1*], and Sophie Guéron[1*]

[1] Laboratoire de Physique des Solides, CNRS, Univ. Paris-Sud, Université Paris Saclay, 91405 Orsay Cedex, France.

[2] Institute of Microelectronics Technology and High Purity Materials, RAS, 6, Academician Ossipyan str., Chernogolovka, Moscow Region, 142432, Russia.

[3] Institut de Chimie Moléculaire et des Matériaux d'Orsay Bâtiments 410/420/430, Univ. Paris-Sud 11, UMR 8182, Rue du doyen Georges Poitou, 91405 Orsay cedex – France.

*Correspondence to: sophie.gueron@u-psud.fr, helene.bouchiat@u-psud.fr

**Abstract**: Spin-orbit interactions are known to have drastic effects on the band structure of heavy-element-based materials. Celebrated examples are the recently identified 3D and 2D topological insulators. In those systems transport takes place at surfaces or along edges, and spin-momentum locking provides protection against (non-magnetic) impurity scattering, favoring spin-polarized ballistic transport. We have used the measurement of the current phase relation of a micrometer-long single crystal bismuth nanowire connected to superconducting electrodes, to demonstrate that transport occurs ballistically along two edges of this high-spin-orbit material. In addition, we show that a magnetic field can induce to 0-π transitions and $\phi_0$-junction behavior, thanks to the extraordinarily high g-factor and spin orbit coupling in this system, providing a way to manipulate the phase of the supercurrent-carrying edge states.

**Main Text:** Reducing the size of a conductor usually decreases its conductivity because of the enhanced effect of disorder in low dimensions, leading to diffusive transport and to weak, or even strong localization. Notable exceptions are the ballistic chiral one-dimensional edge states of the quantum Hall effect, or the recently discovered spin-polarized, counter-propagating edge states of the quantum spin Hall effect (QSHE) found in 2D topological insulators, that are protected from scattering by spin-momentum locking (1,2). Demonstrating such ballistic conduction of one-dimensional states has remained a challenge: most demonstrations have relied on interference experiments (3), non-local measurements in Hall-bar shaped samples (4), or magnetic field-induced interference patterns of the critical current in Superconductor/Topological Insulator/Superconductor (S/TI/S) Josephson junctions (5,6). However a direct signature of the ballistic transport in these materials is still lacking. To probe the transport regime, we use the extreme sensitivity of the relation between the Josephson current $I_J$ flowing through a nanostructure and the superconducting phase difference at its ends $\varphi$, the "current-phase" relation (CPR) (7). It is well known that the CPR of the Superconductor/Insulator/Superconductor (SIS) Josephson junction is sinusoidal $I_J(\varphi)=I_c\sin\varphi$. The CPR of a ballistic Superconductor/Normal metal/Superconductor (SNS) junction, on the other hand, is a characteristic sawtooth in a long junction (for which $L\gg\hbar v_F/\Delta$) (8,9), or segments of a sine in a short junction (Fig. 2C) (10) (L is the N length, $v_F$ the Fermi velocity and $\Delta$ the superconducting gap). Any disorder smooths the CPRs. In this Report, we measure the supercurrent versus phase relation of a Josephson junction made of a three-dimensional monocrystalline bismuth nanowire with topological surfaces connected to two superconducting electrodes. We find a CPR that is the superposition of two sawtooth-shaped signals of slightly different periods, demonstrating one dimensional ballistic conduction along two edges of the wire. Strikingly, the phase of the Josephson current can be shifted away from zero (a realization of the so-called Josephson $\varphi_0$-junction) by a magnetic field, or jump by $\pi$ in a field-controlled 0-$\pi$ transition.

Monocrystalline bismuth nanowires were grown by sputtering, and individual nanowires with (111) top facets were selected using Electron Backscatter Diffraction to exploit the predicted topological edge states of the (111) surfaces (11,12) brought about by the Bi lattice symmetries and high (eV-range) atomic spin-orbit coupling. The wires are narrow enough (thickness and width between 100 and 200 nm) that most normal state conduction is due to surfaces and edges, with only a small contribution of the bulk states (13,14). Furthermore, tight binding simulations of a Bi nanowire with a rhombic section and top and bottom (111) facets (Fig. 1 and (15)) predict one edge state along each (111) facet, extending the result of Murakami (11). These edge states hybridize with the diffusive (100) metallic surface states on the adjacent faces, but nonetheless clearly dominate the local density of states (Fig. 1).

To measure the CPR of the S/bismuth nanowire/S (S/Bi/S) junction, we inserted a single Bi nanowire with (111) surfaces into an asymmetric SQUID setup. The setup consists of a high-critical-current Josephson junction, made of a W superconducting nanoconstriction, in parallel with an S/Bi/S junction characterized beforehand (Fig. 1 and (15)). In this configuration the modulation of the SQUID's critical current by the magnetic flux yields the CPR of the junction with the smallest critical current. While the CPR of atomic point contacts (10), nanowire-based quantum dots (16,17), and graphene (18) have previously been measured in this way, the CPR of micrometer-long, quasi-ballistic channels has, to our knowledge, not been accessed.

The switching current of the asymmetric Bi-SQUID (Figure 2A) clearly displays sawtooth-shaped oscillations of amplitude 400 nA, superimposed on the 80 µA critical current of the high-$I_c$ nanoconstriction. The oscillation period of 9.5 G corresponds to a flux quantum h/2e divided by the SQUID loop area of 2 µm². Those oscillations demonstrate the sawtooth CPR characteristic of a long, perfectly connected ballistic channel - this is a key result of our paper. Eleven harmonics are visible in the Fourier transform of the signal at 100 mK. The 1/n decay of the $\sin(n\varphi)$ harmonics expected for a perfect sawtooth crosses over to a slow exponential decay $\exp(-0.19n)$ (fig 2B). This decay can be interpreted in terms of a quasi-perfect transmission coefficient $t > 0.9$ (15).

Predicted by our numerical simulation, a second path can be identified upon closer inspection of the CPR : a wiggle with a smaller period is superimposed to the main sawtooth signal. The full critical current modulation is reproduced by adding to the main sawtooth the contribution of a second sawtooth with a four times smaller amplitude and a 10% smaller period than the first. This indicates that this second, ballistic, path is situated along the outer wire edge, since the area delimited by the two wires edges is one tenth of the SQUID loop area. Two satellite peaks are visible in the Fourier transform next to the two first main harmonics (inset of fig 2B), whose exponential decay $\exp(-0.45n)$ yields a transmission coefficient $t\sim0.75$. The smaller supercurrent carried by the second channel could be due to a disordered region between the contact and this edge (inset Fig. 1).

The amplitude of the measured supercurrent as well as its resilience to decay at high magnetic fields (15,19) are consistent with the existence of a small total number of channels, each of them confined to an extremely narrow region in real space. The maximum supercurrent through one ballistic channel is $I_1=\pi\Delta/\Phi_0 = e\Delta/2\hbar\sim$ 250 nA for a short junction (i.e. much shorter than $\xi_S= \hbar v_F/\Delta\approx$600 nm). It is smaller for a long junction (20,21), of the order of $ev_F/L\approx$ 100±30 nA for L=1 μm and $v_F = 6\pm2$ $10^5$m/s (22). The critical current of the nanowire, given by the modulated current amplitude of 400 nA (fig 2A), thus implies that three to six perfectly transmitted channels carry the supercurrent. A reasonable assumption is that one path contains three to four quasi perfect channels, each with the same sawtooth-shaped CPR, and all situated at the inner edge of one (111) facet of the wire. They could be associated to the orbitals $p_x,p_y,p_y$ of Bi, as suggested by Murakami (11,12), or could also run along the edges of three parallel terraces at the facet edge. The smaller contribution of the second path is attributed to one or two other channels of smaller transmission, at the outer edge of the other (111) facet (see sketch in Fig. 2B). These findings of two distinct current paths are consistent with the two-path interference pattern measured before integration in the asymmetric SQUID configuration (15). Moreover, the fact that the CPR is visible up to a field $B_{max}\approx0.5$ T, implies that very little flux is threaded through one path, i.e. different channels within a given path must be narrower and closer to one another than $\Phi_0/B_{max}L\sim4$ nm.

The purely one dimensional nature of the Andreev bound states that carry the supercurrent across the nanowire implies that they are insensitive to orbital dephasing and therefore that the magnetic field acts primarily through the Zeeman effect. A Zeeman field can induce a crossing of the Andreev levels, turning an energy maximum into a minimum (23): this causes a sign change of the CPR, or equivalently a π shift of its phase. 0-π transitions are expected when dephasing by the magnetic field equals dephasing by the propagation time through the wire, i.e. when the Zeeman energy equals the Thouless energy, $g_{eff}$ $\mu_B$ B= $\hbar<v_F>/L$. The characteristic field $B_{x,y}\sim$ 600 G, $B_z\sim$ 400 G between 2 successive 0-π transitions (see Fig. 3 that displays the CPR as a function of a magnetic field in the (111) plane, either perpendicular or parallel to the wire axis) yields an effective g factor $g_{eff}\approx$30-100, consistent with the high g factors of some bands in Bi (24). We note that penetration of vortices in the superconducting electrodes would also lead to phase jumps. This is however unlikely as no sign of hysteresis was found in our data. This realization of a 0-π transition induced by the Zeeman field is possible because the junction is long, contains few channels, and the $g_{eff}$ are high enough that the transition occurs at a magnetic field below the superconducting electrodes' (relatively high) critical field.

The usual conversion of phase difference into an electrical (super-)current can be supplemented by a conversion into spin current in the presence of strong spin-orbit interaction. Consequently, a magnetic field can shift the CPR by a phase $\varphi_0$. This effect is predicted to be greatest in a field perpendicular to the wire axis (17,23). Such "$\varphi_0$-junctions" are visible in Fig. 3, in the form of an oscillating phase of the CPR with increasing field, outlined by the winding hatched curve, with a stronger effect in a field perpendicular to the wire (compare Fig 3a to 3b).

In conclusion, the current-phase relation of a Bi-nanowire-based Josephson junction demonstrates that conduction occurs along ballistic channels confined at two edges of the wire's (111) facets. Whereas narrow edge states were predicted for a bilayer of (111) Bi (11,12), it was not obvious that such states should exist in thicker samples, until one-dimensional edge states were detected by scanning tunneling microscopy recently, in two layer-deep pits at the (111) surface of bulk Bi crystals (25). Our work shows that in the superconducting proximity regime, the contribution of the nanowire's two ballistic channels outweighs that of the more numerous diffusive channels on the wire's non topological surfaces. This is because the relative contribution to the supercurrent is proportional to the ratio of Thouless energies, and the diffusive Thouless energy of the surfaces $\hbar D/L^2$, with D the diffusion constant, is more than ten times smaller than the ballistic Thouless energy of the edges $\hbar v_F/L$. Moreover, topological edge states benefit not only from faster ballistic transport, but also from better coupling to the superconducting electrodes due to their spin-moment locking that should produce perfect Andreev retro-reflections (26). By contrast, the slower diffusive carriers also undergo regular (as opposed to Andreev) reflections on opaque barriers, which decrease the supercurrent they can carry. Our results provide a firm foundation on which to base the exploration of the topological character of these edge states.

We acknowledge fruitful discussions with M. Aprili, C. Beenakker, S. Bergeret, D. Carpentier, P. Simon, J. Jobo, S. Bayliss, T. Wakamura, M. Houzet, F. Konschelle, J. Enrique Ortega, A.Yazdani and financial support from CNRS, ANR MASH ANR-12-BS04-0016 and ANR DIRACFORMAG ANR-14-CE32-0003.

Figure captions:

Figure 1: Building an asymmetric SQUID around a (111) Bi nanowire. A: Scanning Electron Micrograph of the bismuth nanowire with (111) top and bottom facets (brown), connected to high-$H_c$ superconducting tungsten electrodes (blue). Dashed rectangle: the S/Bi/S junction prior to integration in the asymmetric SQUID. The supercurrents induced at low temperature through this S/Bi/S junction and others are described in Supplementary Materials. Inset: zoom of the contact area, explaining the asymmetry between the two conduction channels. B: Micrograph of the asymmetric SQUID built after characterizing the S/Bi/S junction: a high-critical-current tungsten nanoconstriction is in parallel with the smaller-critical-current S/Bi/S junction of Fig. 1A. C: Calculated local density of states (LDOS) of an infinitely long, 5 bilayer-thick (111) bismuth nanowire with rhombohedral section.

Figure 2: Sawtooth-shaped CPR of the S/Bi/S junction seen in critical current of asymmetric SQUID. A : Critical current of the asymmetric SQUID, at 130 mK (blue) and 1K (red), revealing the bismuth junction's CPR and combination of two sawtooth-shaped currents $\Sigma(-1)^n/n$ {sin(n$\varphi$)exp(-0.19n)+0.25sin(1.1*n$\varphi$)exp(-0.45n)} (black). Here $\varphi=2\pi B_z S_{int}/\Phi_0$, with $\Phi_0$=h/2e and $S_{int}$ is the area delimited by the squid loop and the inner edge of the wire. B : Fourier transform of the measured curves at 130 mK and 1K. Inset displays the satellite peak on the CPR's Fourier transform due to the second conduction path at the outer edge of the wire (delimiting an area $S_{ext}$), next to the main peak (corresponding to the area $S_{int}$). C : Theoretical zero-temperature current-phase relations of SNS junctions from tight binding simulations in different regimes: Green, short junction with one N site, the CPR is close to the ideal relation $I_s$=2e$\pi\Delta$/h sin($\varphi$/2) sign($\pi-\varphi$); blue, sawtooth CPR of a long ballistic junction, calculated for length L=1.2 $\xi_s$. Purple: rounded CPR of a L=1.2 $\xi_s$-long channel with on-site disorder W/t=0.5, corresponding to a mean free path of L/2 (27). Red, sinusoïdal CPR $I_s$=e$\pi\Delta$/h sin$\varphi$ of a multichannel SIS junction with normal state resistance h/2e². D: Critical current of a reference asymmetric SQUID, made of a similar tungsten constriction in parallel with a superconducting aluminum/oxide/aluminum SIS tunnel junction. As expected, the measured CPR is sinusoidal (dashed line is a sinusoidal fit to the data). The small effect of the circuit inductance has been corrected for. This correction is negligible for the Bi SQUID because of its smaller dimensions.

Figure 3 : Zeeman and orbital effects on CPR: $\varphi_0$-junctions and 0-$\pi$ transitions. A,B : Color-coded Critical current of the asymmetric squid at 100 mK, as a function of both a magnetic field $H_z$ perpendicular to the (111) surface, and in the (111) surface plane, either parallel (A) or perpendicular (B) to the wire axis. The critical current variations are exactly the current-phase relation of the bismuth nanowire. The dashed lines outline how the phase of the current-phase relation shifts with magnetic field ($\varphi_0$

junctions), and in some instances, jumps by π (0- π transitions). Note that a linear relation between $H_x$ ($H_y$) and $H_z$ was subtracted, to account for the small (few percents) projection of the $H_z$ field on the horizontal sample plane. This treatment eliminates the linear variations of the CPR's phase with magnetic field.

C : Sketch of the wire orientation. D: Sections of the color-coded plot as the $B_x$ field causes a 0-π transition.

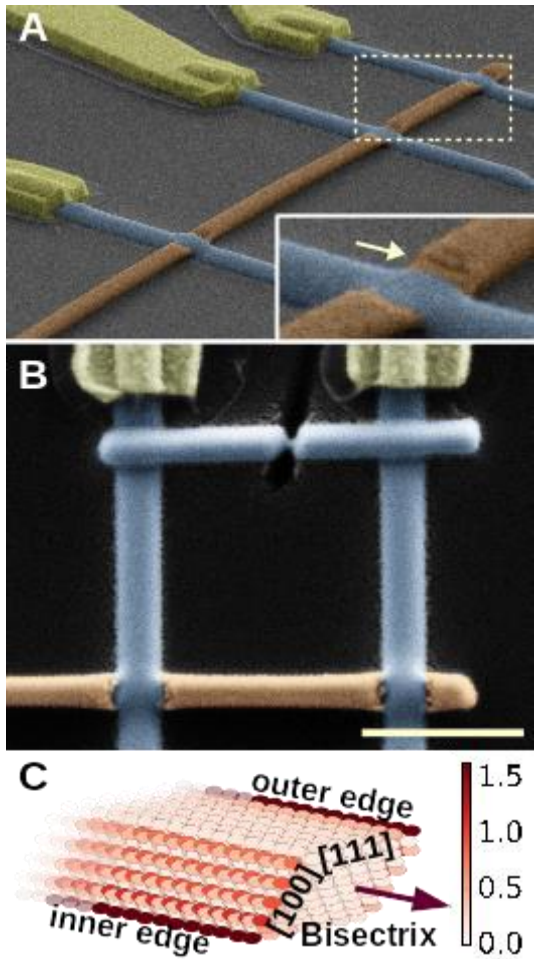

*Figure 1*

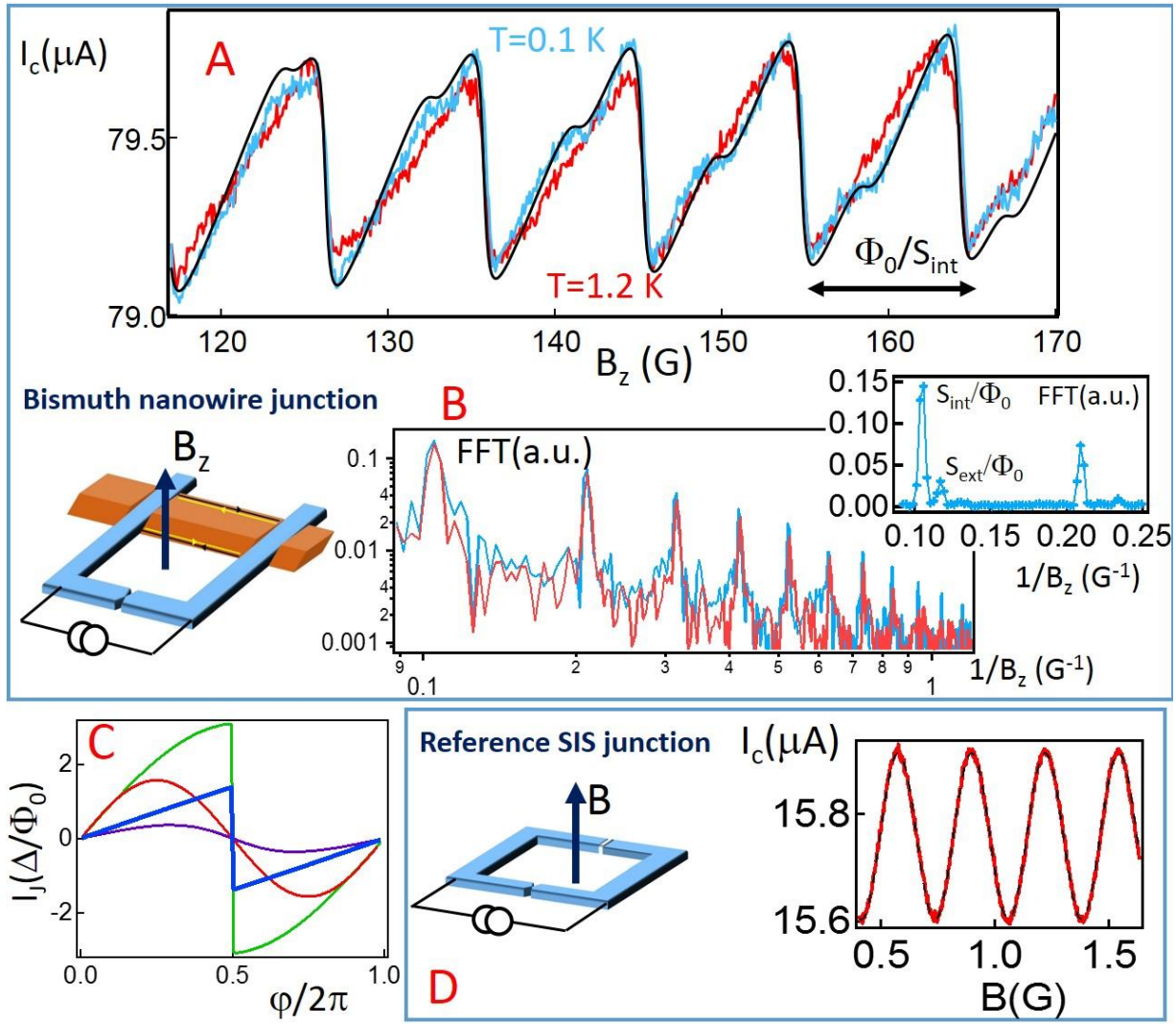

Figure 2

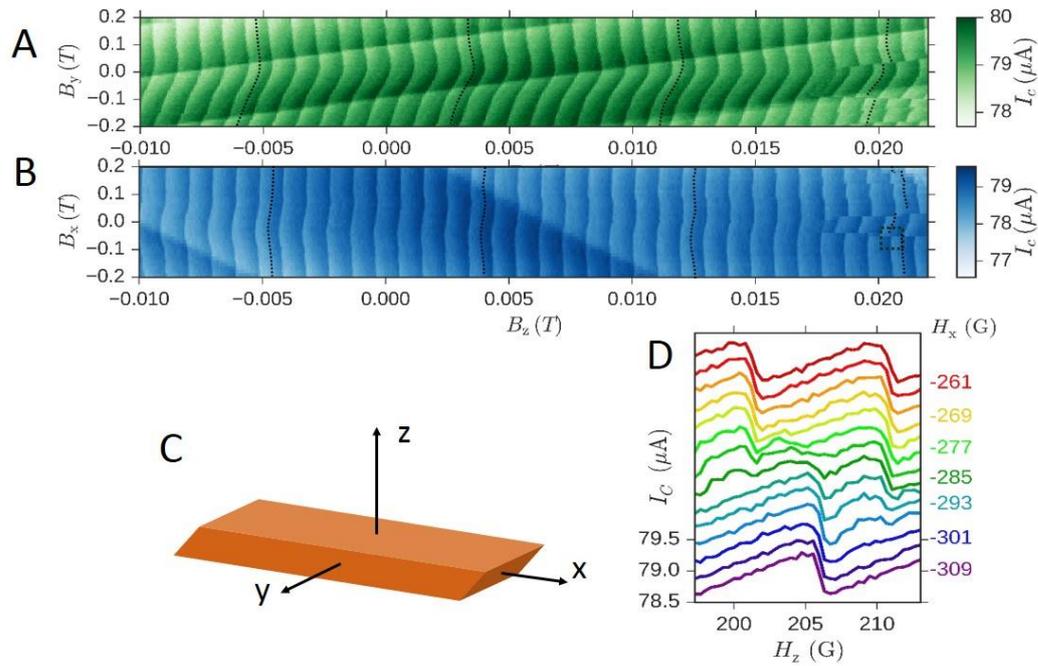

*Figure 3*

# Supplementary materials

Table S1

Fig S1 – S9

References (28 – 35)

S1-Materials and methods;

Bismuth nanowires were grown by RF-sputtering a Bi target of 99.9999% purity onto various substrates at 473 K and in an argon pressure of 10 mBar argon pressure. High resolution TEM observations indicated high quality single crystals, as well as clear facets, with typical radii of 100 − 300 nm.

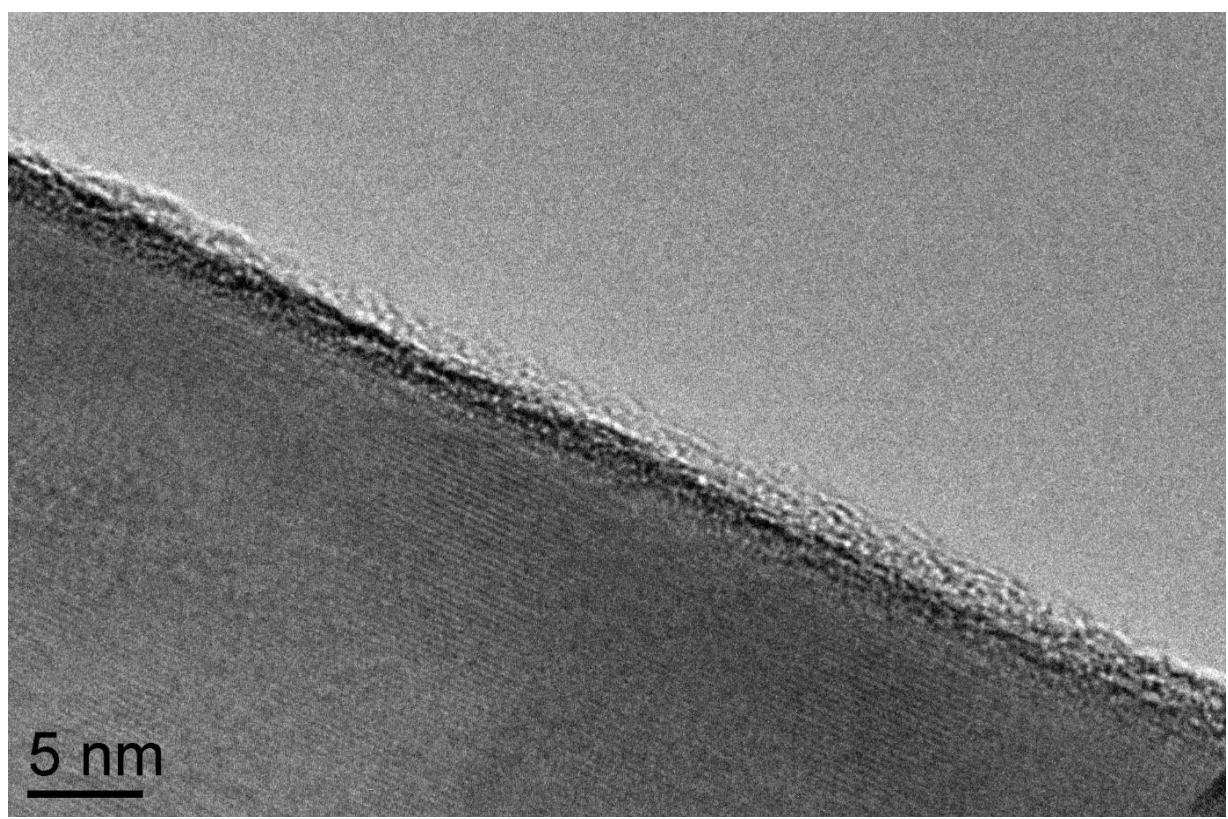

*Figure S4 : High resolution TEM image of a Bismuth nanowire, demonstrating the cristallinity of the wire and the extremely thin oxide layer.*

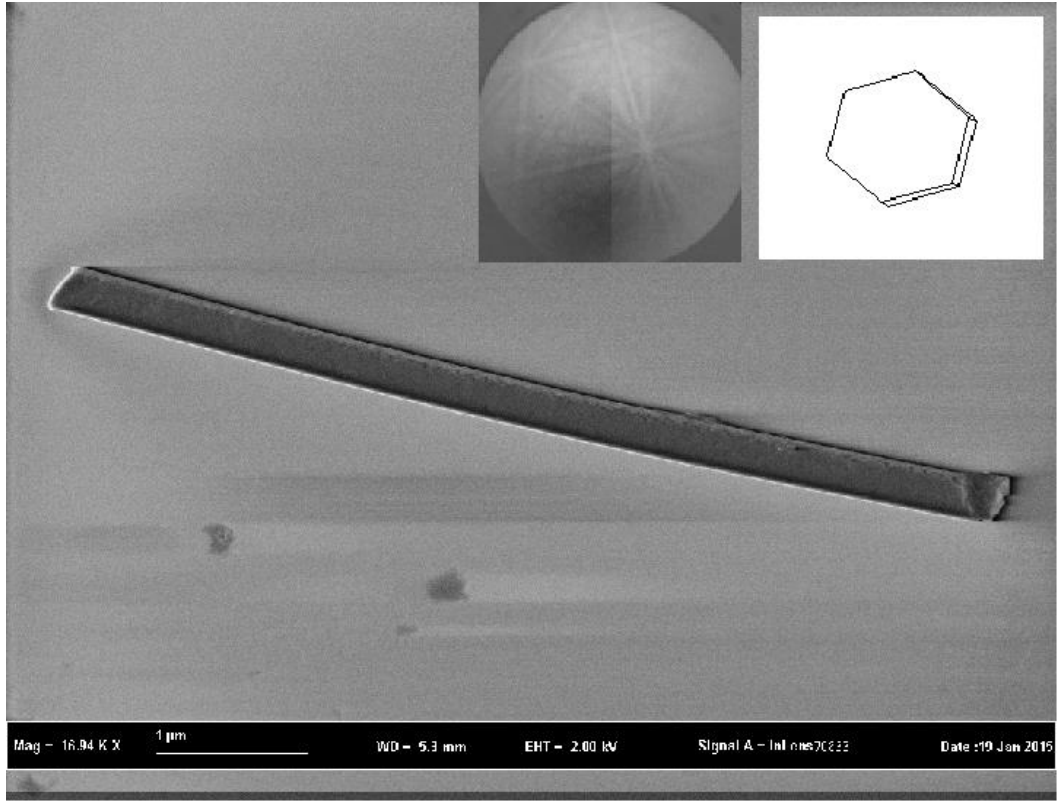

*Figure S2: Scanning Electron micrograph of a Bi nanowire with a top 111 facet, as demonstrated by the Electron Backscattered Diffraction: inset shows the Kikuchi lines pattern (top left) and the reconstructed unit cell (top right).*

The nanowires were then dry deposited on an oxidized silicon substrate. After optical selection, their crystalline orientation was determined using electron backscatter diffraction (EBSD). An example is shown in Fig. S2. This characterization was performed at the extremity of the nanowires to avoid electron beam induced damage of the Bi wire. We also checked on similar nanowires that their crystalline orientation was constant as we changed the position of the beam spot, as is expected for single crystals. At this stage, we selected nanowires whose top surface was determined to be oriented perpendicular to the trigonal [111] axis. Electrical connections were then made using Gallium Focused Ion Beam (FIB)-deposited superconducting tungsten wires (28). For all the measured nanowires, the lengths between the W lines was chosen to be greater than 1 μm in order to avoid possible superconducting contamination. Previous studies using the same setup showed that this could be an issue for wires whose length is below 200 nm. For the same reasons, we minimized the total exposure time of the nanowires under the FIB to a single scan at high scanning rate and low magnification. We connected nine segments with different lengths, from a total of three such nanowires. Their resistances and lengths are summarized in the table S1, and show low contact resistance on average.

Table S1: Characteristics of bismuth samples measured in S/Bi/S two wire configuration. Three different nanowires, labeled $s_1$, $s_2$ and $s_3$, were investigated. Each sample corresponds to a different segment of the nanowire. Length is distance between superconducting contacts, $R_N$ is the normal state resistance at low temperature, determined by the resistance jump above the critical current $I_c$ for the three shortest segments ($s_1$JU $s_2$SD and $s_3$WH). For longer segments that do not have a zero resistance at low bias, $R_N$ is the resistance measured with a bias current of 1 μA. $\tau_{tr}/m^*$ is the transport time divided by the effective mass (in bare electron mass units), deduced from the magnetoresistance's parabolic dispersion. $R_N$ at RT is the sample resistance at room temperature.

| Nanowire | Section | Length (μm) | $R_N$ (Ω) | $\tau_{tr}/m^*$ ( ps ) | $R_N$ at RT |
|---|---|---|---|---|---|
| Bi 1 | $s_1$JU | 1.4 | 330 | 18.6 | 693 |
| | $s_1$UR | 3.0 | 1050 | 17.5 | 1000 |
| | $s_1$RM | 3.5 | 1350 | 21.9 | 1100 |
| Bi 2 | $s_2$SD | 1.7 | 275 | 19.0 | 478 |
| | $s_2$DX | 3.5 | 400 | 18.0 | 605 |
| | $s_2$Xa | 4.5 | 450 | 19.1 | 643 |
| | $s_2$aZ | 4.1 | 390 | 18.9 | 623 |
| Bi 3 | $s_3$KW | 3.8 | 218 | 22.4 | 488 |
| | $s_3$WH | 2.4 | 150 | 22.8 | 500 |

The samples were then cooled to 100 mK and their critical current Vs magnetic field was measured using a lock-in detection technique.

After this first characterization step, the Current Phase Relation (CPR) was measured on the JU segment using the asymmetric SQUID technique (10) with a reference junction made of a W constriction. To this end, a FIB-deposited W wire was added between the two tungsten electrodes, in parallel to the Bi nanowire (Fig. 1), and was subsequently etched with the Ga+ beam while measuring the total resistance between the contacts, until the total resistance reached 190 Ω, corresponding to a constriction resistance of 300 Ω. The critical current of the SQUID was deduced from the average of 100 to 400 measurements of the switching current, using a counter synchronized to a current ramp at 180 Hz, and triggered by the jump of the sample resistance. In those measurements, the sample resistance was measured at 100 kHz with a lock-in detector operating with a time constant of 1 ms.

Figure S3 displays the normal state resistance of each section of the three measured nanowires (described in Table S1), as a function of their length. It displays a linear behavior, characteristic of a diffusive regime. Stars are measured resistance, solid line is linear fit with the constraint that the contact resistance is positive. The measurements were carried out at 4.2 K, except for section s1JU. The resistance of that section was deduced from the resistance jump at the switching transition, which underestimates the resistance by the wire/superconductor contact resistance. Thus that point was not included for the fit. The slopes yield a resistance per unit length of 358 Ω, 57 Ω and 50 Ω/μm for $Bi_1$ $Bi_2$ and $Bi_3$ respectively. The dispersion between these values could be due to variations in wire diameters (which may vary by up to a factor of two).

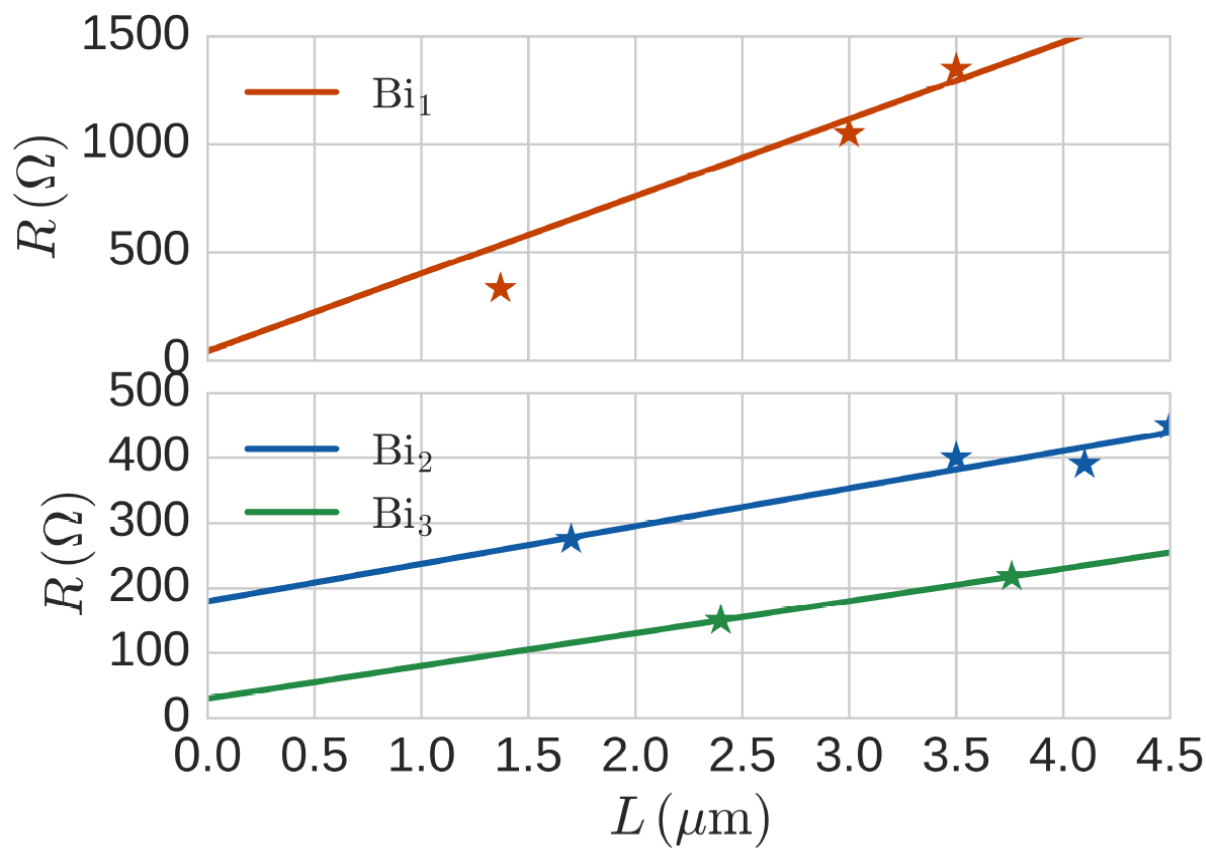

*Figure S3: Normal state resistance of each section of the three measured nanowires (table S1), as a function of their length*

S2- Tight binding simulation

Tight binding simulations were performed on a 1D infinite geometry corresponding to the stacking of 5 (111) Bi bilayers, 7 atoms wide, arranged to provide a simplified model of the rhombic section of the nanowire studied in this paper. The z direction therefore corresponds to the (111) direction or trigonal axis in Bi's rhombohedral lattice, which was determined experimentally by electron diffraction (EBSD). The infinite direction of the nanowire and the direction of the lateral surface were chosen in order to match two criteria: they should be of high symmetry since nanowire growth mechanisms favor high symmetry surface orientation, and should reproduce the rhombic section that was inferred by SEM observation. The retained geometry corresponds to the (100) lateral surface and a growth direction along the bisectrix axis. This corresponds to the case studied by Murakami in the 1 bilayer thickness limit and 20 atoms wide.

We find a sharp density of states along two edges, that correspond to zigzag type A edges (defined in (**Erreur ! Source du renvoi introuvable.**)) : those are the edges whose top atoms are less coupled to the bulk.

The Hamiltonian matrix was written according to the sp3 tight binding model of Liu and Allen (33) using the KWANT python package (34). The local density of states at the Fermi level was computed by summing the local spectral weights in a window of 0.1 eV around the Fermi level. A smaller window would yield even sharper density of states on the zigzag type A edges.

S3-Relative contributions from bulk and surfaces in the normal state

Transport in the normal state is due to bulk and surface states (in addition to topological edge states), and we now characterize the properties of both. ARPES experiments (13) yield a surface carrier density $n_s \approx 10^{17}/m^2$ and a corresponding Fermi wavelength $\lambda_s \approx 8$ nm. The effective mass is $m_s \approx 0.2\, m_e$ where $m_e$ is the free electron mass. The band structure of bulk Bi is quite complex, leading to several types of carriers (13). In a nanowire, quantum confinement prevents the presence of light electrons whose Fermi wavelength is greater than the wire's transverse dimensions. Thus, it is likely that the majority bulk states are hole states with an effective mass $m_b \approx 0.065\, m_e$ and a Fermi wavelength of $\lambda_b \approx 60$ nm. Those numbers suggest that in a 1 micron-long wire of width 200 nm and height 100 nm, the two non-topological surfaces contain roughly 100 times more carriers than the bulk.

Thus we can assume that the greatest contribution to normal state conductance comes from the surfaces, $G_N = G_s + G_b \sim G_s$. The mean free path on the surfaces is then deduced from the 300 Ω resistance of a 1 micron by 200 nm surface, via $\sigma_S = G_Q k_F l_s = GL/W = 1.7\, 10^{-2}\, \Omega^{-1}$, from which we deduce a surface-state mean free path $l_s \approx 300$ nm, and transport time $\tau_s \approx 3$ ps.

Interestingly, whereas the bulk states do not contribute much to the zero field conductance, we find that they yield an important contribution to the low field magnetoresistivity $\delta\rho(B)$ because their mobility $\mu_b \approx e\tau_b/m_V$ is greater than the mobility of the surface carriers, due to their smaller effective mass. Whereas elongated wires are not expected to exhibit sizable low field magnetoresistance, the situation is different when the conductivity $\sigma = Ne\mu$ results from 2 types of carriers with different mobilities (31):

$$\frac{\delta\rho(B)}{\rho} = B^2(\mu_b - \mu_s)^2 \frac{\sigma_b \sigma_s}{\sigma^2} \quad (S1)$$

Since the surface conductivity is greater than the bulk conductivity eq (S1) becomes

$$\frac{\delta\rho(B)}{\rho} = e^2 B^2 \left(\frac{\tau_b}{m_b} - \frac{\tau_s}{m_s}\right)^2 \frac{\tau_b}{\tau_s} \frac{N_b}{N_s} \frac{m_s}{m_b} .$$

The magnetoresistivity coefficient, which is nearly sample independent, yields

$$\frac{\delta\rho(B)}{\rho} = (2\ ps)^2 \cdot \left(\frac{e}{m_e}\right)^2 \cdot B^2$$

So that we find $\tau_b \approx 2$ ps, and $l_b \approx 200$ nm. The bulk mobility $\mu_b$ is thus 2 times greater than the mobility of the surface states $\mu_S$.

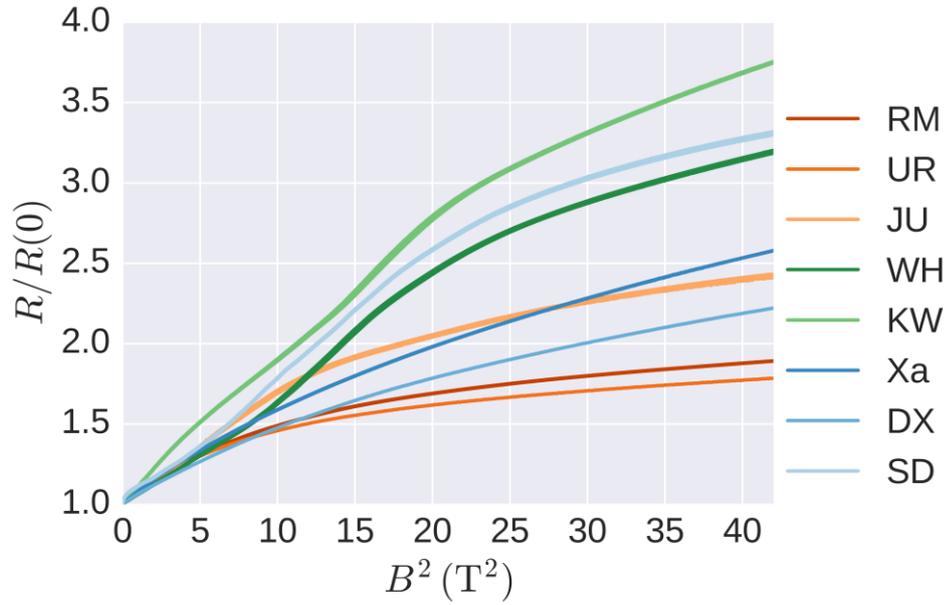

Figure S4: Magnetoresistances of each section normalized by the zero field resistance and plotted against the square of the magnetic field $B_z^2$. A quadratic dependence on magnetic field is clearly visible on all samples at low magnetic fields, allowing the determination of the transport times given in table S1.

S4- Critical current measurements of other (111) Bi nanowires and other sections of this nanowire in two contact configuration, before integration in an asymmetric squid, low and high field dependence.

We have measured several sections of three different (111) nanowires. To induce superconductivity, we connect them to high-critical-field-superconducting electrodes one to a few micrometers apart (see Table S1). This leads to a sizable supercurrent at subKelvin temperature for three sections of those wires, whose length is of the order of or smaller than 2μm (Table S1). As shown in Fig. S5 for two samples $s_1$JU and $s_3$WH, we find that the critical current does not decay with magnetic field over the expected small range of one flux quantum through the wire's surface area, as in multichannel diffusive samples (29,30). Instead, the supercurrent persists up to unusually high fields, corresponding to several hundreds of flux quanta, persisting up to 10 Tesla in some samples. This is a clear indication of minimal orbital dephasing within Andreev pairs by the magnetic field, that is only possible if very few narrow conduction paths shuttle Andreev pairs across the wire, as suggested in previous experiments on Bi nanowires of unknown orientation (19). For comparison, we also show in Fig. S5 how the supercurrent induced in an Ag nanowire with the same aspect ratio as the Bi nanowire, connected to the same type of superconducting tungsten electrodes, is suppressed by a magnetic field above just a few $\Phi_0/S$, where S is the wire area perpendicular to the field direction.

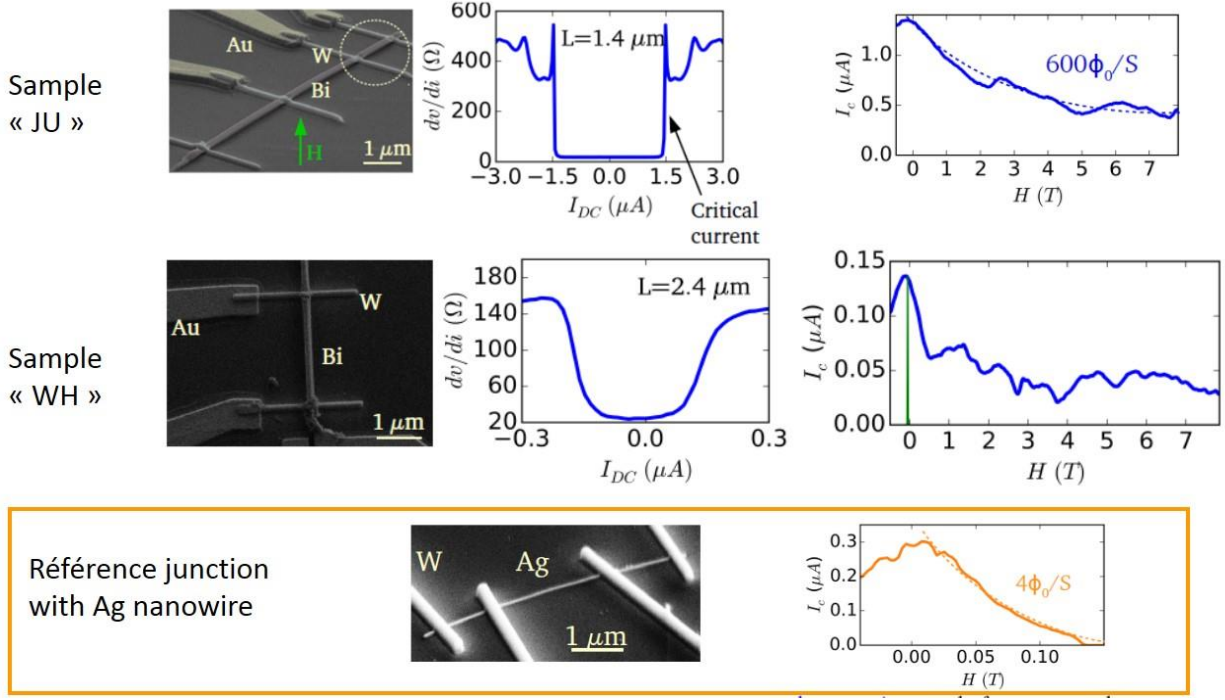

*Figure S5 : Differential resistance versus current and field dependence of the critical current of two Bi nanowires and one Ag nanowire of similar aspect ratio, connected to the same type of superconducting tungsten electrodes. The critical current of the Bi nanowires extends beyond several hundred flux quanta threading the nanowires, whereas the critical current of the Ag nanowire decays with a typical scale of a few flux quanta threading the wire. This points to many channel diffusive transport in the Ag nanowire, and very few narrow channels in the Bi nanowires.*

The two wire measurement of these Bi nanowires sections also reveal strong field modulations of the critical current, similar to our previous findings in Bi wires (with unknown crystalline orientations), that are the signature of a strong lateral confinement of carriers. Certain samples, including the one (sample "Bi$_1$JU") we have integrated in the asymmetric SQUID, also display a striking (but small) periodic modulation of the critical current by a magnetic field perpendicular to the sample axis, as in a SQUID. The small modulation amplitude suggests that the supercurrent branches into very few highly transmitted, quasi ballistic 1D channels, and into some less well transmitted channels, whose distribution on the bismuth facets can be inferred by tracking the modulation with magnetic field orientation with respect to the nanowire. The period of the modulation is 100 G, close to one flux quantum divided by the projected area of a wire facet on a normal to the field, demonstrates that the interfering channels are situated at the edges of facets. They give rise to the periodic modulation observed in the current phase relation measured when the nanowire is incorporated into an asymmetric squid configuration. We note that the critical current of the Bi nanowire is about 2 times larger than the current measured in the asymmetric SQUID setup described in the main part of the paper. This change in the value of the critical current can be attributed to thermal cycling of the sample. It may also be a signature of a topological crossing at $\phi=\pi$, that would lead to a factor 2 higher critical current measured in a current biased setup compared to the critical current measured in a phase- biased set up (21). New experiments are needed to assess this point.

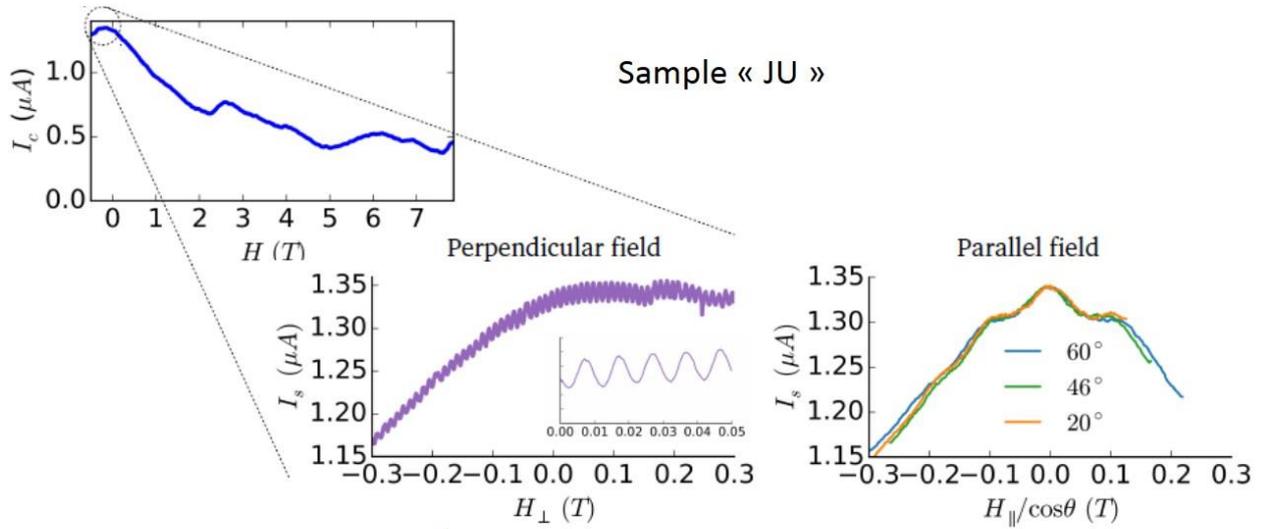

*Figure S6 Modulation of the critical current of Sample $s_1JU$ connected in two-wire configuration, before integration in asymmetric SQUID configuration. Both in-plane and out-of-plane magnetic field generate oscillations of the critical current, with periods of 100 G for a field along $H_z$ and 800 G for a field along the y axis, qualitatively consistent with two (unequally) transmitted edge states at the "A" edges of the Bismuth nanowire (see fig 2). Right panel: Different curves with field applied at different angles in the (x,y) plane overlap when the field value is renormalized by the cosine of its angle with respect to the y axis.*

S5- Self induction effects

It is well known that self induction effects can strongly distort the current phase relation in SQUIDS (7,35). In our experiments these effects were minimized by making the SQUID loops small enough. Screening of the flux into the loop of a SQUID is characterized by the parameter $\beta= 2\pi LI_c/\phi_0$ where L is the inductance of the branch containing the small junction. This parameter is estimated to be respectively 0.1 and 0.06 for the DC SQUIDs containing the SIS and Bi nanowire-based Josephson junctions. These values of β yield small corrections on the SIS SQUID, and negligible ones for the Bi-based SQUID.

In practice, we have evaluated β by adding to the geometrical inductance the (much greater) kinetic inductance of the W wire. The kinetic inductance is related to the normal state resistance $R_N$ via $L_K= R_Nh/2\pi\Delta$ (32), yielding 30 pH per micrometer of the FIB-deposited W wires, and 1 pH per micrometer of Al wire. The inductance used of the SIS reference SQUID thus amounts to 200 pH, yielding β=0.1.

S6-Attenuation of the sawtooth: Thouless energy from temperature dependence

From the harmonics content of the current phase relation one can give a lower bound of the transmission of the wire. The harmonics of the critical current of a long wire of transmission *t* in the N state are expected to decay as $t^{2n}$ for large n (the exponent 2 comes from the transmission of Andreev pairs: $t^2$ instead of t for quasi-particles). From the measured decay of the harmonics ~ exp(-0.19n), we can deduce that t > 0.9. Other effects, such as high frequency noise, may also contribute to this decay.

As expected, the decay is more pronounced at T=1 K, and is proportional to exp-0.23n ~ exp-n(T/$E_{Th}$)) leading to a small rounding of the sawtooth. The decay factor gives a lower bound estimate of the Thouless energy, $E_{Th}$≈ 5 K, from which one gets a Fermi velocity of 6 $10^5$m/s.

### S7- Simulations: Critical current amplitude, from short to long junction

Figure S7 displays the critical current amplitude of ballistic SNS junctions of varying lengths, from the short to the long limit, using a tight binding calculation described in (27). In the short junction limit L<$\xi_s$ we find $I_J = \Delta /\Phi_0$ and in the long junction limit the Josephson current is $ev_F/L$ where $v_F = 4\pi ta$ the Fermi velocity and $\xi_s = 2ta/\Delta$ are expressed in terms of the lattice spacing a and the hoping energy t. The red line corresponds to the long junction limit $ev_F/L$.

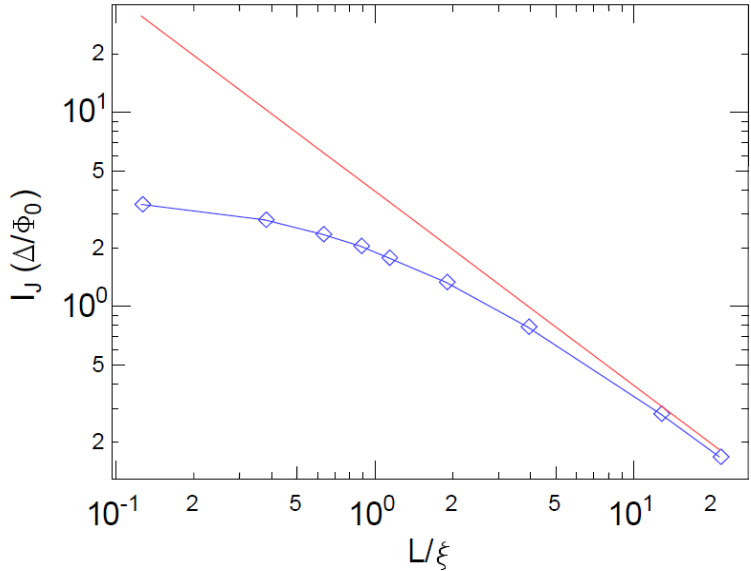

*Figure S7 : Calculated critical current amplitude $I_J$, in units of $\Delta/\Phi_0$, from short to long SNS junction. L is the length of the N segment.*

### S8: SQUID-like oscillations of the critical current for a 2D topological insulator

We have also performed simulations on the hexagonal lattice with second neighbor spin-orbit interactions following the Kane and Mele model for the quantum spin Hall state. As expected when the Fermi energy lies in the spin-orbit gap, the Andreev spectrum of a ribbon connected to superconducting reservoirs consists of 2 spin-degenerate levels, crossing at the Fermi energy for a phase difference of π. They correspond to the counter-propagating 1D states along the edges of the ribbon. The amplitude of the Josephson current is found to oscillate with the flux through the ribbon with an amplitude which depends on the relative transmission at the NS interfaces along these 2 edges. This is shown in Fig. S8 below, and compared to oscillation amplitude deduced from data in Fig.2.

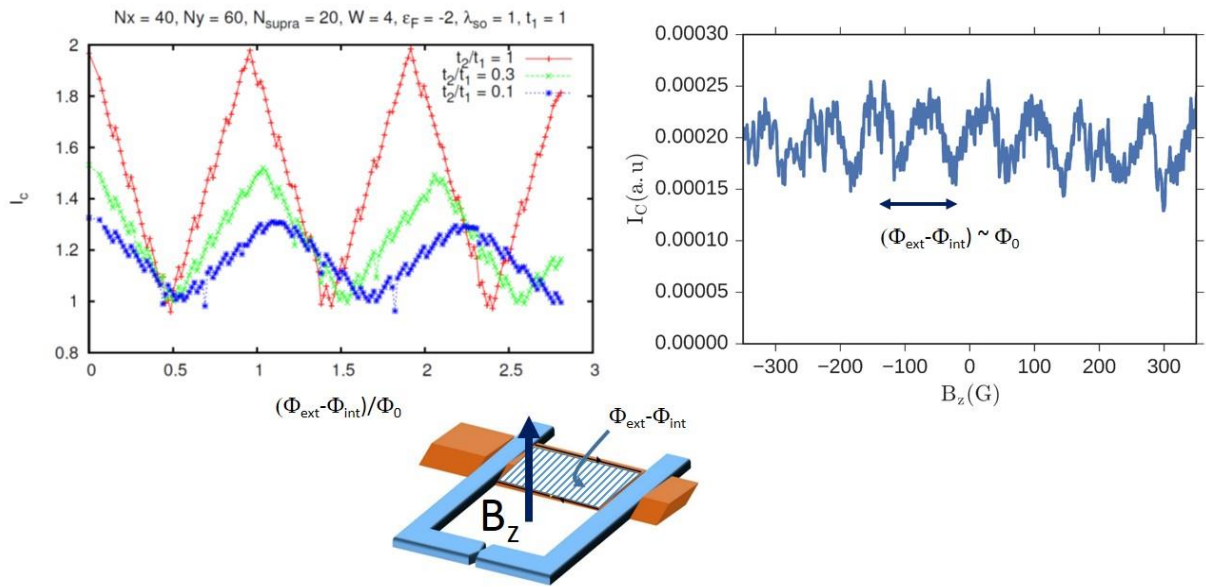

*Figure S8 : Oscillation of the current amplitude with a periodicity given by a flux quantum in the nanowire area. Comparison between theory for two topological edge states with different transmissions t1 and t2, and the modulation amplitude extracted from the experiment.*

S9: Second asymmetric SQUID circuit with a Bi nanowire as a weak link

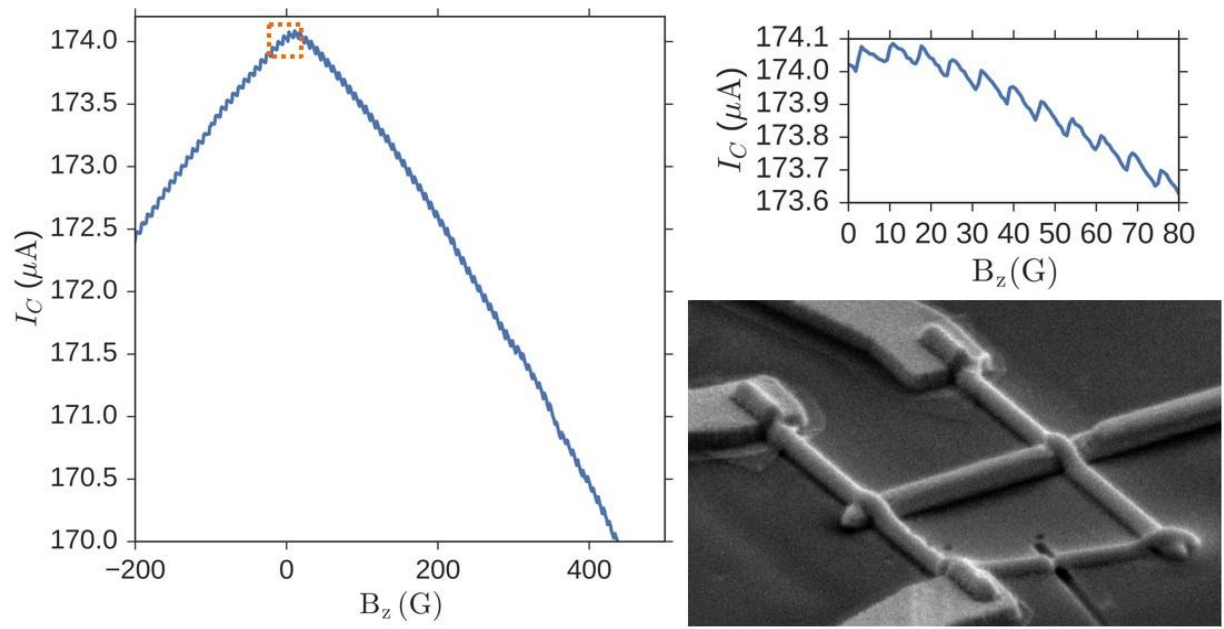

*Figure S9 Critical current of a second asymmetric SQUID, revealing the sawtooth CPR of a second Bi(111) nanowire.*

As shown in Fig. S9, the sawtooth shaped CPR was reproduced in a second sample made with the same technique as the one shown in the main text. The asymmetry in this device is extremely large: the critical current of the W constriction was 173 µA and the modulation amplitude was 45 nA. Nevertheless, a periodic sawtooth signal in magnetic field with period of 8G (corresponding to one $\Phi_0$ in the area of the loop) is clearly superimposed on top of a slowly varying background. The second path cannot be identified in the CPR. This could be due to the difficulty to detect it in this extremely asymmetrical squid configuration, or more likely because the lower channel is not connected to the tungsten electrode, in contrast to the sample presented in the main text. The SEM image indeed suggests that the W electrode may not contact the bottom nanowire facet.